\input epsf.tex
\documentstyle[prl,aps,epsf,multicol,rotate]{revtex}
\tighten
\begin{document}
\title{
Statics and dynamics of a one-dimensional quantum many-body
system}
\author{
Eugene B. Kolomeisky$^{1}$ and Joseph P. Straley$^{2}$}
\address{
$^{1}$Department of Physics, University of Virginia, Charlottesville, Virginia 22901 \\
$^{2}$Department of Physics and Astronomy, University of Kentucky,Lexington, Kentucky 40506-0055 }
\maketitle
\begin{abstract}
The macroscopic zero-temperature behavior of weakly-
incommensurate systems in one dimension is described in terms of 
solitons.  The soliton density n obeys equations displaying 
several types of singular interface-like solutions:  (i) 
equilibrium or moving boundary between the n = 0 and finite n 
regions, and (ii) stationary or moving annihilation front 
separating solitons from antisolitons.
\end{abstract}
\vspace{-0.5cm}
\begin{multicols}{2}
\narrowtext
Low-dimensional materials provide a testing ground for the theories
of strongly correlated systems\cite{NATO}.  Recently there have been
several attempts to manufacture truly one-dimensional
systems with periodically modulated parameters.
Kouwenhoven {\it et al.}\cite{kou} built a 
one-dimensional wire of spinless electrons with a 
200 nm periodic potential.  Upon varying the
electron density with a gate voltage they observed
suppression of conductance at some rational fillings.
Recently Tarucha {\it et al.}\cite{Tarucha} has 
succeeded in introducing a shorter period potential.  A different
path was undertaken by van Oudenaarden {\it et al.}\cite{van}
who modelled a one-dimensional quantum liquid placed in a periodic
potential by a very long and narrow array of small Josephson
junctions placed in a magnetic field.  The field determines
the density of
Josephson vortices in the array; these behave like
quantum particles.  A suppression
of the array resistance at some rational fillings was
observed, and 
interpreted in terms of a Mott transition.
Another relevant experimental system would be a weakly-doped nanotube at low
temperature. 
 
In each of these examples, the fundamental model is a one-dimensional
quantum liquid -- spinless fermions, or bosons with a short-ranged
repulsive interaction --
at zero temperature, placed in an external periodic 
potential\cite{Kolo}.
The possible phases are classified as follows:
 
*When the interactions between particles 
are not strongly repulsive, quantum fluctuations will render the 
periodic potential irrelevant, and the system will always be a 
conductor.

*Strongly repulsive particles will become immobilized 
even by a weak commensurate potential, and the resulting state 
has long-ranged order, so that we may idealize this structure as 
being a crystal of particles in registry with the periodic 
potential.  This state is called a Mott insulator.   
 
*When the
number of particles per period of the potential is not rational, 
there can be no commensuration and the system is again a 
conductor.  The macroscopic behavior of such {\it weakly-incommensurate 
}systems is the subject of this paper.
Excluding special cases (such as the "filled band" of one 
particle per potential well), the Mott state is degenerate.  When 
the number of particles per potential well is $\nu  = p/q$, where p
and q are co-primes, there are q different registries of the
crystal relative to the potential.  An important configuration is
the soliton, which is the lowest-energy way of joining two
domains of different registry.  Although there are q-1 such
combinations, we will assume that energetic considerations select
a preferred sequence of neighboring domains, so that the
possibilities reduce to "soliton" and "antisoliton".  Then for $\nu $
close to $p/q$ the structure may be regarded as a low-density gas
of solitons.
This is a weakly interacting system, even though the
original particles are dense and their interactions are not
small.

We note that in the path integral representation the gas of
quantum solitons is equivalent to a gas of classical lines in two dimensions,
and thus describes
the behavior of magnetic flux lines in a
large area Josephson junction (i.e. a thin normal interface between two
superconductors), or steps on the surface of a crystal.

Moving a soliton through the system shifts the Mott crystal by 
one potential well, and thus at $\nu  = p/q$ moving q solitons shifts
p particles past a given point, returning the crystal to its
original registry: a soliton is a fractional particle\cite{heeger}.

Solitons repel each other with a short-ranged force; this can be 
approximated\cite{Pokro} by treating the solitons as a gas of 
noninteracting free fermions, with the chemical potential

\begin{equation}
\mu  = \epsilon  + \pi ^{2}\hbar ^{2}n^{2}/2m
\end{equation}
where $\epsilon $ is the energy cost to introduce a soliton, n is the 
density of solitons, and m is the soliton mass.  The last term is 
a quantum effect coming from the overlap of soliton 
wavefunctions, and is more important than the classical effect 
coming from the overlap of soliton strain fields when the 
intersoliton distance $n^{-1}$ exceeds the soliton width $\xi $.

In equilibrium the chemical potential must be constant along the 
system, i.e. $\epsilon  + \pi ^{2}\hbar ^{2}n^{2}/2m =$ const.  
Depending on the magnitude 
of the constant this equation can have a solution with $n$ real (the
uniform soliton conductor), or may have no real solution at all.
The latter means there are no solitons in the system (Mott state,
$n = 0)$ and the condition of equilibrium is irrelevant as there is
nothing to equilibrate.  The Mott crystal plays a role of a
vacuum of solitons/antisolitons.

When the system is placed in an external field which provides a 
potential energy U(x) for the solitons, the condition of 
equilibrium becomes\cite{LDL}
\begin{equation}
\mu  + U(x) = \epsilon  + \pi ^{2}\hbar ^{2}n^{2}/2m + U(x) = const                  
\end{equation}
In contrast to the translationally-invariant case U = const, this 
equation allows solutions in which a region occupied by solitons 
can coexist with the soliton-free vacuum.

For example, let us look at a system of finite length $0 \le  x \le  L$
placed in a field that exerts a constant force F on each soliton.
If the original particles are charged, then solitons will be
charged as well\cite{Kolo} so that the practical realization of
such field would be the electric field.  Equilibrium in such a
field is only possible if the system is disconnected from
external leads which we will assume to be the case.  Then $U(x) =
- Fx$, and (2) has a solution of the form $n^{2} = (2mF/\pi ^{2}\hbar ^{2})(x - 
x_{s})$.
For $F > 0$ there is a soliton vacuum for $x < x_{s}$, and
the solitons will occupy the $x \ge  x_{s}$ half-space with a sharp
boundary between the two regimes located at $x = x_{s}$.  However,
solitons and antisolitons are oppositely charged, so that the
same constant force
will push them in opposite directions, and there can be a second
region for $x < x_{a}$ where there is a finite density of
antisolitons.  We will associate $n > 0$ with a density of solitons
(an excess of the original particles) and $n < 0$ with antisolitons
(a particle deficit).  Then the equilibrium distribution is
\begin{equation}
n(x) = \left\{\matrix{\matrix{ \sqrt{2mF(x-x_{s})/\pi ^{2}\hbar 
^{2}}&\qquad x > x_{s}&\cr
-\sqrt{2mF(x_{a}-x)/\pi ^{2}\hbar ^{2} }&\qquad x < x_{a}&} }\right.
\end{equation}
The values of x$_{s}$ and $x_{a}$ are constrained by the condition of
conservation of particles, which is also a constraint on the
integral of n(x).

(i)  Assume that for F = 0 the ground state is the Mott crystal 
with no solitons present.  A nonzero field F will promote the 
creation of soliton-antisoliton pairs.  The change of the system 
energy upon creation of a soliton-antisoliton pair is given by 
$\epsilon (y) = 2\epsilon  - Fy$ where the first term is the energy cost  to create
two solitons of opposite kind separated by a distance $y \gg  \xi $ while
the second term is the energy gain in external field.  The pairs
for which $\epsilon (y) \le  0$, i.e. those separated by a  distance bigger
than $2\epsilon /F$ will be present in equilibrium.  As a result the
external field F polarizes the Mott insulator by creating and
spatially separating solitons and antisolitons.  If
the field F pushes solitons to the right and antisolitons to the
left, the density distribution n(x) is odd in x about $x =
0$, and we have n = 0 for $0 < x < \epsilon /F$ and $n = (2mF/\pi ^{2}\hbar 
^{2})^{1/2}(x -
\epsilon /F)^{1/2}$ for $x \ge  \epsilon /F:$ a strip of Mott phase of width 
$2\epsilon /F$
separates solitons from antisolitons.  This conclusion is only
true for a sufficiently large system whose size L exceeds the
size of the Mott strip $2\epsilon /F$; otherwise pair creation is not
profitable and the vacuum is the lowest energy state.  As the
field increases, the Mott strip shrinks, and our description
fails at very large fields of order $\epsilon /\xi $ when the size of soliton-
free region becomes comparable with the soliton width $\xi $.

(ii)  Assume that for F = 0 the ground state is a soliton 
conductor.  A sufficiently small nonzero field F pushing solitons 
to the right will turn the uniform soliton distribution into $n =
(2mF/\pi ^{2}\hbar ^{2})^{1/2}(x - x_{s})^{1/2}$ illustrated in Figure 1a (for this 
case
$x_{s}$ is outside of the physical region).  As the field increases,
at some F there will be a marginal configuration $(x_{s} = 0) ($shown
in Figure 1b) where the soliton density vanishes at the left end
of the system. At larger fields $(x_{s} > 0)$ a soliton-free Mott
strip forms at the left end of the system (Figure 1c).  Upon
further increase of the field when solitons get pushed
sufficiently far away from the left, there will be another
marginal configuration for which the size of the Mott strip is
exactly $2\epsilon /$F. At largest fields, antisolitons nucleate at the left
end of the system $(x_{a} > 0)$, the number of solitons at the right
increase by the same amount, and the size of the Mott strip stays
equal to $2\epsilon /F$ thereafter (Figure 1d).
 
\centerline{
\epsfxsize3.0in
\epsfbox{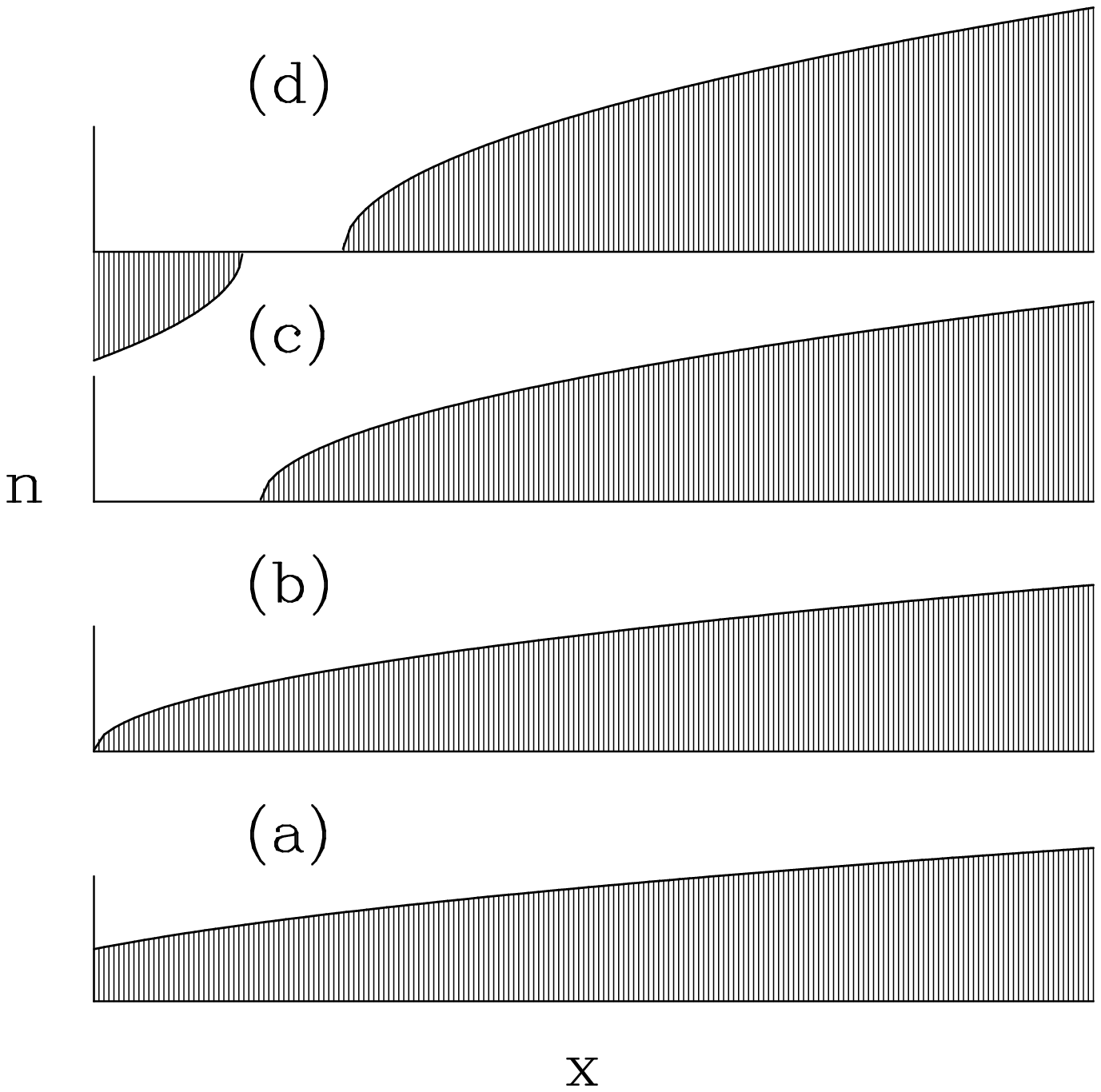}
}
{\it
Soliton distribution in a soliton conductor for various applied 
fields. (a) For small fields, the soliton distribution is 
nonuniform but everywhere nonzero. (b) There is a critical value 
of the field, where the soliton density is zero at one end of the 
system; (c) for larger fields, there is a region where $n(x) = 0.
$
(d) For even larger fields, soliton-antisoliton pairs nucleate.
}

The equilibrium configurations of solitons in an external field 
are related to the equilibrium crystal shapes of three-
dimensional crystals, since the latter system can be viewed as 
equilibrium of atomic steps\cite{Noz}.

We now turn to a discussion of nonequilibrium effects, and 
hereafter we assume that there are no external fields present.  
The force exerted on a given soliton by its neighbors is $- \partial \mu /\partial 
x$,
which will cause it to drift with the velocity $v = -\gamma \partial \mu /\partial 
x$, where
$\gamma $ is phenomenological friction constant.  The resulting current
of solitons is
\begin{equation}
j = nv = -\gamma n\partial \mu /\partial x = -\Gamma \partial \mu 
/\partial x = -bn^{2}\partial n/\partial x
\end{equation}
where $\Gamma  = \gamma n$ is the system mobility, $b = \gamma \pi ^{2}\hbar ^{2}/m$ 
is a dynamical
parameter, and we used Eq.(1) to compute $\partial \mu /\partial $x.  The mobility 
$\Gamma $
is linear in the soliton density n, which implies that the
conductivity of the system vanishes linearly in deviation from
commensuration as the Mott insulator phase is approached from the
conductor side.  This transport theory 
explains the linear drops in resistance seen by van Oudenaarden
{\it et al.}\cite{van} in the vicinity of the Mott phases,
and parallels the flux line
mechanism of resistivity of a type-II superconductor in the
vortex state\cite{Abri}, and the growth regime of vicinal
crystal surfaces via the  motion of steps\cite{Noz}.

Conservation of soliton number within a region implies a 
continuity equation, which provides an equation of motion for 
$n(x,t)$:
\begin{equation}
\partial n/\partial t = - \partial j/\partial x = b {\partial \over 
\partial x} (n^{2} {\partial n\over \partial x})
\end{equation}
When the function $n(x,t)$ changes sign (both solitons and
antisolitons present) the relaxation within the two regions
continues to be described by Eq.(5); however, the behavior at the
interface n = 0 where the solitons and antisolitons annihilate
requires further discussion.

When a soliton-antisoliton pair annihilates, an energy $2\epsilon $ is
released [first term of (1)].  Viewed classically, the force that
corresponds to this potential energy has only the range of order
$\xi $, since solitons do not communicate beyond this scale.  From our
macroscopic viewpoint this would imply no interaction between
solitons of opposite kind at all, and the dynamics of
annihilation would be determined by the rate at which they get
pushed together by the excess of particles elsewhere.  This
overlooks a quantum effect: there always is a finite annihilation
probability of a pair whose size is bigger than $\xi $.  This
possibility of future annihilations gives an effective longer-
range attraction between solitons of opposite type.  However, the
dominant effect associated with the region where solitons and
antisolitons annihilate is that the mobility vanishes [Eq.(4)],
so that attractive forces between solitons and antisolitons have
no important macroscopic effect on dynamics, and (5) is still
valid.

Equation (5) resembles the diffusion equation, but the "diffusion 
constant$" D = bn^{2}$ is density-dependent.  This equation has
previously appeared in the theories of shock waves,
filtration\cite{Zel}, and dynamics of crystal
surfaces\cite{LV}.

The most remarkable property of (5) is that it allows self-
similar moving solutions with a sharp front where the soliton 
density vanishes.  Indeed, assume there is such a boundary.  
During a sufficiently small time interval the velocity of the 
front v can be taken approximately constant.  The distribution of 
soliton density near the front can be sought in the form $n(x -
vt)$.  Substituting this into (5) and integrating twice between
the front position $x_{f} [n(x_{f}) = 0]$ and x we find
\begin{equation}
vn^{2}/2 - j_{f}n + (j_{f}/v)\ln|(vn + j_{f})/j_{f}| = v^{2}(x_{f} - x)/b
\end{equation}
where j$_{f}$ has a meaning of the current density at the front 
position $x = x_{f}$.  The shape of the density profile near the front
edge depends on whether $j_{f}$ is zero or finite:

(i)  The case $j_{f} = 0$ has been considered previously\cite{Zel}.
Taking in (6) the limit $j_{f} = 0$ we find $n^{2} = (2v/b)(x_{f} - x)$ which
implies that the solitons are present in the region of space
satisfying $v(x_{f} - x) \ge  0$.  The region $v(x_{f} - x) < 0$ is soliton-
free which  is consistent with the assumption that the current
density $j_{f}$ vanishes at the front position.  The density
distribution near the front is given by\cite{Zel}
\begin{equation}
n = \pm  \sqrt{{2v(x_{f}-x)\over b}}
\end{equation}
This type of front describes a cloud of solitons/antisolitons 
invading a Mott phase.  As in the equilibrium case (3), the front 
has a square-root singularity (7).  The steepness of the moving 
front is determined by the velocity v of the front.

(ii)  If $n(x_{f}) = 0$ is the annihilation front then solitons and
antisolitons get pushed by their neighbors towards $x = x_{f}$, and
the current density $j_{f}$ must stay finite there.  As far as we can
tell, this type of shock has not been studied previously.  Solving
(6) to
lowest nonvanishing order in $|vn/j_{f}| \ll  1$ the density distribution
near the front can be found as
\begin{equation}
n = [3j_{f}(x_{f} - x)/b]^{1/3} \{1 + (v/4j_{f})[3j_{f}(x_{f} - x)/b]^{1/3}\}
\end{equation}
This type of front describes solitons invading an antisoliton 
conductor (or the other way around).  The peculiar feature of (8) 
is that the magnitude of the force acting on a soliton $\partial \mu /\partial x 
\propto 
n\partial n/\partial x \propto  (x_{f} - x)^{-1/3} \propto  n{ } ^{-1}$diverges at 
the annihilation edge.
This divergence offsets the vanishing of the system mobility and
thus overcomes a dynamical bottleneck.  As a result the current
density (4) stays finite at the annihilation edge which is the
ultimate reason for the leading cubic-root singularity in (8).

As Eq.(8) shows, there can be both moving and stationary 
annihilation fronts.  In fact, for $v = 0, $ Eq.(8) provides the
exact {\it steady-state} solution to (5).  The parameter $j_{f}$ will now
have the meaning of a fixed current flowing through the system
which is the same everywhere.

The front coordinate x$_{f}$, velocity $v = dx_{f}/dt$,  and annihilation
current $j_{f}$ entering Eqs.(6)-(8) are undetermined functions of
time which should be found from the complete solution of the
problem.  For the special case of a constant velocity front, we
have $x_{f} = vt +$ const, and (7) becomes an exact solution to the
problem of a constant velocity soliton invasion of the Mott
phase.  Similarly, Eq.(6) provides an exact solution to the
problem of a constant velocity annihilation front which can be
visualized as follows:

Assume v $>$ 0 so that the front moves to the right.  Far to the 
left of the front position, $x \ll  x_{f}$, the front profile  is
approximately given by (7).  Far to the right, $x \gg  x_{f}$, the
density tends to a constant value $n_{\infty } = - j_{f}/v$, and then the
solution (6) can be approximated by $n = n_{\infty }\{ 1 - \exp[(v/bn_{\infty 
}^{2})(x_{f} -
x)]\} $.  This is exactly the density profile one would get ahead of
a constant velocity front if the relaxation could be described by
the linearized version of (5) with the diffusion constant $D =
bn_{\infty }^{2}$; the size of the perturbed region ahead of the front is
given by the diffusion length $D/v = bn_{\infty }^{2}/v$.

Below we consider a few more practically relevant examples when 
the front velocities are not necessarily constants.

(i)  Imagine a Mott insulator for which extra particles have been 
added to a region of the system.  This can be described as a 
segment of nonzero soliton density imbedded between two half-
infinite commensurate domains.  The number of particles added 
determines the total number of solitons N, which is conserved: 
$\int^{\infty }_{-\infty }n(x,t)dx =$ N.   
Arbitary initial particle
distributions will relax asymptotically to a common
form\cite{Zel}
\begin{equation}
n(x,t) = \pmatrix{{N^{2}\over \pi ^{2}bt}}^{1/4} \sqrt{1-\pi 
x^{2}/4N\sqrt{bt}}
\end{equation}
which has the characteristic square-root singularity (7) at its 
edges.

(ii)  Consider the case that the average density of particles is 
appropriate for commensuration but the initial distribution is 
inhomogeneous.  For example, we could have a Mott insulator 
everywhere except in a small region, where there is an excess of 
particles on one side, and a deficit on the other.  In the 
soliton description, the two regions will contain solitons of 
different types --solitons and antisolitons -- which can be 
described by a single function n(x).  Here are two examples in 
which soliton-antisoliton annihilation plays a role:

Consider a periodic distribution of solitons of both signs such 
that the total soliton "charge" is zero.  The spatial periodicity 
of the distribution will be preserved by time evolution; this 
implies that the asymptotic solution to Eq.(5) should be sought 
in the form\cite{LV} $ n = t^{-\alpha }f(x)$.  It is readily seen that the
equation of motion (5) determines the exponent to be $\alpha  = 1/2$.
The density decays because solitons succeed in getting through
the zero mobility region, by having a singularity in the applied
force: the function f(x) has $\Delta x^{1/3}$ singularities at every point
where it changes sign\cite{LV} in agreement with the general
argument leading to (8).

A related example is a soliton distribution such that $n(-x) = -
n(x)$. This symmetry will be preserved by the time evolution.  Due
to annihilation events at $x = 0$, the total number of solitons of
either type is not conserved;  however, the dipole moment
$P = \int^{\infty }_{-\infty }xn(x,t)dx$ is conserved\cite{Zel} by the  equation of
motion (5).  The expansion
rate is smaller than in previous case of the spread of a region
of excess solitons because of annihilation of soliton-antisoliton
pairs at $x = 0$; the envelope decays more slowly than in the
periodic case because the solitons are not confined.  The
analytic solution for this case is\cite{Zel}
\begin{equation}
n(x,t) = {AP\over x_{f}} \pmatrix{{x\over  
x_{f}}}^{1/3}\left[\matrix{^{B}{ } ^{-}\pmatrix{{x\over x_{f}}}^{4/3} 
}\right]^{1/2}
\end{equation}
where A and B are constants. At the extremal edges, there is 
again the $\sqrt{\Delta x}$ singularity, (7), but near the origin $n(x) \propto  x{ } 
^{1/3}
$
[see Eq.(8)], as is required to have a net current of solitons
through the mobility bottleneck.

\end{multicols}
\end{document}